\documentclass[twocolumn,aps,showpacs]{revtex4}

\usepackage{amsmath}
\usepackage{bm}              

\usepackage{graphicx,amssymb}

\begin{document}

\title{{\it Ab initio} calculation of structural and electronic properties of
                  Al$_x$Ga$_{1-x}$N and In$_x$Ga$_{1-x}$N alloys}
\author{E. L\'opez--Apreza$^1$, J. Arriaga$^1$, and D. Olgu\'\i n$^2$}
\affiliation{%
  \textsuperscript{1}\,Instituto de F\'{\i}sica, Universidad Aut\'onoma 
de Puebla, Apartado Postal J--48,72570, Puebla, M\'exico\\
  \textsuperscript{2}\,Departamento de F\'{\i}sica, 
Centro de Investigaci\'on y de Estudios Avanzados del 
Instituto Polit\'ecnico Nacional, A. P.
14740, 07300, M\'exico D.F., M\'exico}

\begin{abstract}
Using  the  density functional theory (DFT) with  the  generalized
gradient   approximation  (GGA),  the  structural  and  electronic
properties  of  wurtzite AlN, GaN, InN, and their related  alloys,
Al$_x$Ga$_{1-x}$N  and  In$_x$Ga$_{1-x}$N,  were  calculated.  
We  have  performed
accurate {\it ab initio}  total energy calculations  using  the  
full--potential  linearized  augmented plane wave  (FP--LAPW)  method  to
investigate  the  structural and electronic  properties.  
In  both
alloys  we  found that the fundamental parameters  do  not  follow
Vegard's law.  
The lattice parameters, $a,\ c,$ and $u$, for the 
Al$_x$Ga$_{1-x}$N alloy are found to exhibit downward bowing, 
while for In$_x$Ga$_{1-x}$N
there  is  an  upward  bowing for the $a$ and  $c$  parameters  and  a
downward  bowing  for the internal parameter, $u$.  
Furthermore,  we
found  that  for both alloys, the band gap value does  not  follow
Vegard's  law.  
As a by--product of our electronic  band  structure
calculations, the effective masses of the binary compounds as well
as  their  related  alloys  were  calculated.  
We  show  that  the
calculated properties for the binary compounds, as well as for the
studied  alloys, show good agreement with most of  the  previously
reported results.  
Finally, using the frozen phonon approach,  the
A$_1(TO)$  mode  for the different systems studied in this  work  was
calculated. 
Our calculations show good agreement with experimental
values  reported for the binary compounds. 
For the ternary alloys,
our  calculations reproduce experimental values for  Al$_x$Ga$_{1-x}$N  as
well as theoretical predictions for In$_x$Ga$_{1-x}$N.
\end{abstract}

\maketitle

\section{Introduction}

The  III--Nitride semiconductors have attracted much attention over
recent   years   because  of  their  potential   applications   in
technological  devices. This is due mainly to the  fact  that  the
energy  gap  can  be  tuned over a wide spectral  range  from  the
visible to the ultraviolet regime of the electromagnetic spectrum.
Although  the zinc--blende and wurtzite structures are  present  in
the  GaN,  AlN,  and InN semiconductors, it has been  demonstrated
experimentally  that wurtzite is the most stable structural  phase
of  these  compounds.  Moreover, due to their  high  chemical  and
thermal  stability,  the  III--Nitrides are  ideal  candidates  for
applications  under  extreme conditions such as  high  temperature
applications. In the wurtzite crystalline structure the  value  of
the  band gap ranges from 0.8 eV for InN \cite{Davydov}, 3.4 eV for GaN
\cite{YKoide} and 6.2 eV for AlN \cite{Morkoc}, providing a huge interval of
energies for this parameter whenever the concentration forming the
alloy  is carefully selected. The hexagonal wurtzite structure  is
extensively utilized because all of the III--nitride semiconductors
and  their related alloys exhibit a direct band gap energy,  which
results  in  a  high  emitting  performance \cite{4,5}.  Due  to   the
remarkable  progress in epitaxial growth technology, high  quality
samples  of these compounds can be produced. High--quality wurtzite
InN is currently available and its direct band gap energy has been
determined  to  be around 0.8 eV, which is much smaller  than  the
commonly  accepted value of 1.9 eV \cite{Davydov}. From  a  theoretical
point of view, many calculations using different methods have been
done  to  characterize  the structural,  electronic,  and  optical
properties  of these systems; however there is still no  agreement
in  the  scientific  community concerning the  values  of  certain
parameters, since they show significant scattering when we compare
the experimental or theoretical results published in literature,
as we discuss in this work.

In  this paper, by means of numerical calculations based on  first
principles,  we  present a study of the structural and  electronic
properties  of the AlN, GaN, InN semiconductors and their  related
alloys,  Al$_x$Ga$_{1-x}$N and In$_x$Ga$_{1-x}$N, with a wurtzite  
structure.  The
analysis  was  made  by calculating the total  energy  from  first
principles. First, we analyzed the binary compounds, GaN, AlN, and
InN,  and then their related alloys, Al$_x$Ga$_{1-x}$N and 
In$_x$Ga$_{1-x}$N,  for
$x=0,\ 0.25,\ 0.50,\ 0.75,\ {\rm and}\ 1.0$. 
Our calculations were based on the
density  functional  theory (DFT) using the  generalized  gradient
approximation (GGA) in order to calculate the exchange--correlation
term  in  the total energy. We used the Wien2k simulation  package
developed  by the Vienna University of Technology. In section  II,
we  describe the model used in this work, while in section III, we
discuss  our results and compare them with data found  in  related
literature. Finally, in section IV, we present our conclusions.

\section{Theoretical Aspects and Computational Method}

Our  calculations  were  performed within  the  framework  of  the
density functional theory (DFT) \cite{Hohenberg}, which states that all
of  the ground state properties of a system are functionals of the
electron density and the total energy is expressed in terms of the
electron density rather than the wave function. At present, DFT is
one  of the most accurate methods to calculate the structural  and
electronic  properties of solids. We have used the  full--potential
linearized augmented plane wave method (FP--LAPW) as implemented in
the Wien2k code \cite{PBlaha}. As most of the first principles methods,
LAPW  is  a procedure used to solve the Khon--Sham set of equations
for  the  density of the ground state, the total energy,  and  the
eigenvalues  of  a many--electron system. In the present  analysis,
the exchange--correlation energy of the electrons was treated using
the generalized gradient approximation (GGA) \cite{Perdew}. To minimize
the  energy,  the  Wien2k code divides the  unit  cell  into  
non--overlapping  spheres  centered at atomic  sites  (muffin--tin  (MT)
spheres), and the interstitial region. In the MT spheres, the 
Khon--Sham orbitals are expanded as a linear product of radial functions
and  spherical  harmonics, and as a plane wave  expansion  in  the
interstitial  region.   The basis set inside  each  MT  sphere  is
divided into core and valence subsets. The core states are treated
within the spherical part of the potential only and are assumed to
have  a  spherically symmetric charge density that  is  completely
confined within the MT spheres. In this work, the valence part was
treated within a potential that was expanded into harmonics up  to
$l=10$. We have used MT sphere radii of 1.6 Bohr for N, 1.9 Bohr for
Al,  2.0  Bohr  for Ga, and 2.33 Bohr for In. The  self--consistent
calculation  was considered to converge when the total  energy  of
the  system was stable to within $10^{-5}$ Ry. Care was  taken  to
assure  the  convergence  of the total  energy  in  terms  of  the
variational cutoff energy parameter. Furthermore, we have used  an
appropriate  set  of  k--points to compute  the  total  energy.  To
calculate  the convergence of the total energy we wrote the  basis
functions up to a cutoff radius of $R_{mt}K_{max} = 7$ Ry for both of  the
binary  compounds and the Al$_x$Ga$_{1-x}$N alloy, and R$_{mt}K_{max} = 8$ 
Ry  for
the  In$_x$Ga$_{1-x}$N  alloy. We have minimized the  total  energy  using
different  sets  of  k--points  in  the  irreducible  part  of  the
Brillouin  zone and constructing an appropriate grid in  the  unit
cell  according  to the Monkhorst--Pack procedure \cite{Monkhorst}.  The
number  of k--points used was chosen in order to assure convergence
within our accuracy criterion ($10^{-5}$ Ry).

In  the wurtzite structure, the positions of the atoms inside  the
unit  cell are $(0,0,0)$ and $(2/3, 1/3, 1/2)$ for the cation Al,  Ga,
In; and  $(0, 0, u)$ and $(2/3, 1/3, 1/2+u)$ for the anion N, where  
$u$ is  the  internal  parameter for the cation--anion  separation.  
We began  our study by optimizing the structural parameters  for  the
binary  compounds,  GaN,  AlN, and InN, starting  from  the  ideal
wurtzite  structure  with a ratio $c/a=1.633$ and  $u=0.375$  for  the
internal  parameter. This optimization was made  by  an  iterative
process  as  a function of the volume, $V$, the $c/a$ ratio,  and  the
internal parameter, $u$, until the total energy converged to  within
0.01 mRy. To model the Al$_x$Ga$_{1-x}$N and In$_x$Ga$_{1-x}$N 
ternary alloys,  we
used  a 32--atom supercell with periodic boundary conditions.  This
corresponds  to a $2\times2\times2$ supercell which is twice the size  
of  the
primitive  wurtzite unit cell in both directions: along the  basal
plane  and along the $c-$axis. We minimized the total energy for different
values  of  the  concentration, $x$ (0.25, 0.50, and 0.75), as a
function  of  the  three  variables mentioned  above. The atomic
electronic  configuration  used in our calculations was:  
Al (Ne, 3p, 3s), Ga (Ar, 3d, 4s, 4p), In (Kr, 4d, 5s, 5p), N (He, 2s, 2p). 
The  Ga3d  and
In4d  electrons are treated as valence band states using the local
orbital extension of the LAPW method \cite{PBlaha}.

\section{Results and Discussion}

\subsection{Structural parameters for the Al$_x$Ga$_{1-x}$N and 
In$_x$Ga$_{1-x}$N alloys}

\subsubsection{Al$_x$Ga$_{1-x}$N}

\begin{figure*}[t]%
\includegraphics*[width=\linewidth]{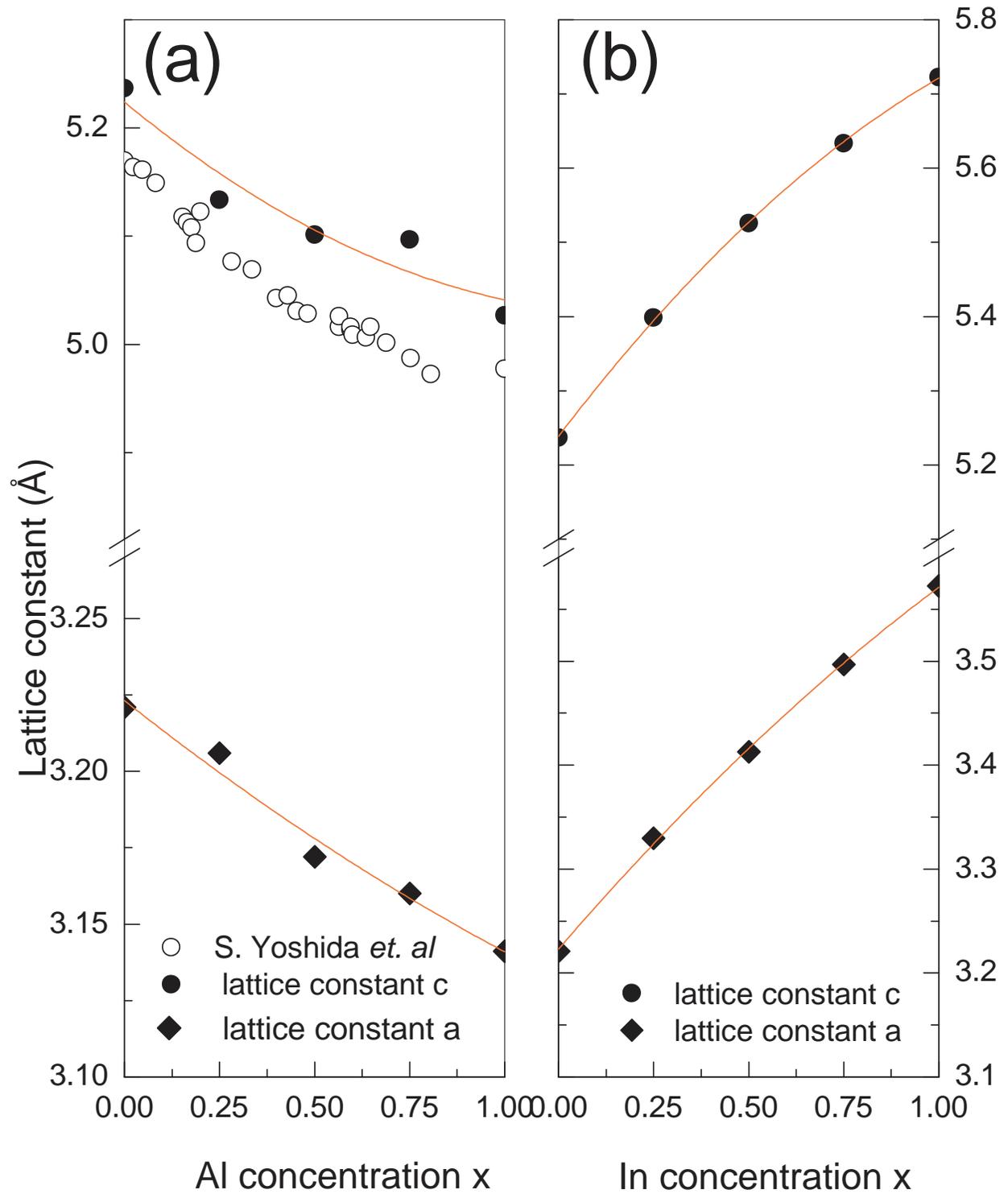}
\vspace{-2cm}
\caption{(a) Lattice constants $a$ and $c$ for the
Al$_x$Ga$_{1-x}$N alloy as a function of the
Aluminium composition, $x$. Open circles show the experimental results
from Ref \cite{SYoshida}. (b) Lattice constants $a$ and $c$ for the
In$_x$Ga$_{1-x}$N alloy as a function of the
Indium composition, $x$.} 
\label{figura1}
\end{figure*}

Table 1 summarizes our results and compares them with experimental
and   theoretical  results  reported  previously  using  different
methods.  In going from GaN to AlN, when the Al--content increases,
the  values  of  the  lattice parameters of  the  Al$_x$Ga$_{1-x}$N  alloy
decrease. This is due to the fact that the size of the Al atom  is
smaller  than  the Ga atom. This is not the case for the  internal
parameter, $u$,  in which we observe an increase of this  parameter
when  we  increase the Al content of the alloy. Figure  \ref{figura1}a
depicts  the  behavior of the $a$ and $c$ parameters as a function  of
the  Aluminum concentration. We can see from the figure that these
parameters show a clear deviation from the linear behavior  stated
by  Vegard's law, which determines the parameter behavior of these
alloys  with a zinc--blende structure \cite{BTLiou}. The deviation  from
Vegard's  law can be quantified by adjusting the curves in  Figure
\ref{figura1}a to the following formula:
\begin{equation}
A(x) = x A_{\rm AlN} + (1-x) A_{\rm GaN} - \epsilon_A x(1-x);
\label{uno}
\end{equation}
where  $A(x)$ stands for the different structural parameters, $a,\  c,$
and  $u$,  of  the  Al$_x$Ga$_{1-x}$N  alloy. 
$A_{\rm AlN}$  ($A_{\rm GaN}$)  represents  the
structural parameters of the binary AlN (GaN) compound and  $\epsilon_A$  
is the  respective  bowing  parameter for the  lattice  and  internal
parameters. If we fit our calculated values from Table \ref{tabla1} to
Eq. (\ref{uno}), we obtain a set of bowing parameters: 
$\epsilon_a = 0.016$ \AA, 
$\epsilon_c  =0.119$  \AA,  and  $\epsilon_u = 0.002$. 
We observe that all bowing  parameters
have  a  positive value, which indicates a downward bowing,  being
the  lattice constant $c$ which possesses a larger deviation  from
the  linear  Vegard's law, as was experimentally  reported  by  S.
Yoshida  et al. \cite{SYoshida}, and by Yun et al. \cite{FYun}. Other  works
report  that the lattice parameters follow Vegard's law  (see  for
example  Angerer  et  al.  \cite{HAngerer});  however,  from   from   a
theoretical approach, this alloy has only been studied  using  the
virtual crystal approximation (VCA) by M. Goano et al. \cite{MGoano} or
by  using  first  principles by Z. Dridi et  al.  \cite{ZDridi}.  Other
authors  have  reported  a  non linear  behavior  of  the  lattice
parameters,  but their work demonstrates an upward bowing  of  the
parameters \cite{Liou}.

To   compare   our   results   with  the  available   experimental
information,  we plot in Fig. \ref{figura1}a the calculated  value  of
the  $c$  parameter  obtained  by  Eq.  (\ref{uno})  together  with  the
experimental  results found in literature \cite{SYoshida}.  As  we  can
see,  our  calculated values show the same trend depicted  by  the
experimental data. Other examples of experimental studies for this
alloy  can  be  found  in  the works by D.  K.  Wickenden  et  al.
\cite{DKWickenden}, and by K. Itoh et al. \cite{KItoh}.

\subsubsection{In$_x$Ga$_{1-x}$N}

The   recent   developments  in  blue--green  optoelectronics   are
essentially  due  to the high efficiency luminescence  of  
In$_x$Ga$_{1-x}$N/GaN   heterostructures.  
Despite  their   importance,   several
properties  of  In$_x$Ga$_{1-x}$N  alloys are not  fully  understood.  
For example,  the optical properties of InN crystals are poorly  known
since  the  available  growth  techniques  have  not  allowed  the
production  of high quality epitaxial layers. Recent  improvements
in  the  molecular beam epitaxy (MBE) technique have  led  to  the
availability   of   high  quality  InN  films.   Photoluminescence
measurements of these films indicate an energy gap around 1 eV  or
less \cite{Davydov}. The In$_x$Ga$_{1-x}$N alloy has been 
studied theoretically
by  several  groups  using  different methods.  M.  Goano  et  al.
\cite{MGoano}  used  pseudopotentials to compute the  gap  through  the
virtual  crystal approximation approach. Z. Dridi et al.  \cite{ZDridi}
used  LDA  FP--LAPW and the virtual crystal approximation approach.
J.  Serrano et al. \cite{JSerrano} worked within the framework  of  the
density   functional   theory  (DFT)  with   the   local   density
approximation (LDA) using the Ceperley--Alder form for the exchange--
correlation energy. C. Stampfl et al. \cite{CStampfl} utilized the DFT,
the LDA, and the GGA of Perdew et al. for the exchange--correlation
functional.  A.  Zoroddu et al. \cite{AZoroddu} from  first  principles
within  the  DFT utilized the plane--wave ultrasoft pseudopotential
method  within  both the LDA and the GGA. And P.  Carrier  et  al.
\cite{PCarrier}  used plane--wave pseudopotentials and the  LAPW  method
with the LDA.

\begin{figure*}[t]%
\includegraphics*[width=\linewidth]{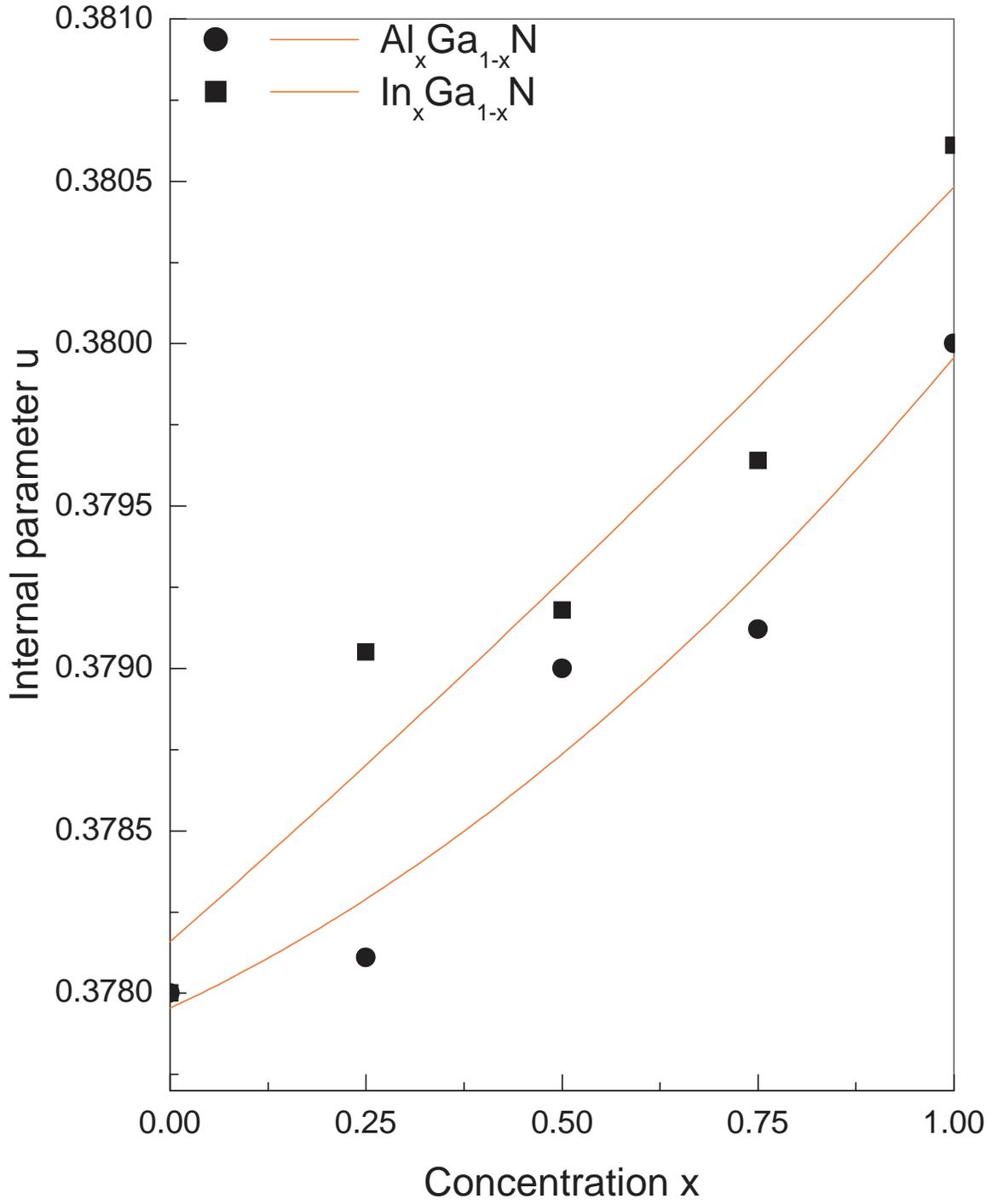}
\vspace{-2cm}
\caption{Internal parameter $u$ for the
Al$_x$Ga$_{1-x}$N and
In$_x$Ga$_{1-x}$N alloys a function of the
Aluminium and Indium composition, $x$, respectively.}
\label{figura2}
\end{figure*}

In  Table \ref{tabla2} we summarize our results and compare them with some of
the  representative theoretical and experimental results found  in
literature. We can see from these results that the values of the $a$
and $c$ parameters increase when the In concentration increases.  We
plot these results in Fig. \ref{figura1}b. It is clear from the figure
that, as in the previous case, there is not a linear dependence in
these two parameters with an increase in the In concentration.  If
we adjust these results using Eq. (\ref{uno}), as we did for the 
Al$_x$Ga$_{1-x}$N  alloy,  we  obtain: $\epsilon_a = -0.140$ \AA, 
$\epsilon_c = -0.188$  \AA,  and  $\epsilon_u  =0.0001$.   
For  this  alloy  we  can  observe  that  the  deviation
parameters  for both $a$ and $c$ have a negative value, which  implies
an  upward  bowing  and is clearly observed in Fig.  \ref{figura1}b.
This  is not the case for the internal parameter, $u$, which  has  a
nearly linear dependence with the In concentration as demonstrated
by  the  bowing  parameter, $\epsilon_u = 0.0001$.  The  simulation  results
indicate that the $c$ lattice constant has a larger deviation from
the  linear  Vegard's law when compared with the lattice  constant
$a$. Finally in Fig. \ref{figura2}, we plot the internal parameter, $u$, for
both  alloys as a function of the concentration, $x$. Solid  circles
and  squares  correspond to our theoretical results for  Al$_x$Ga$_{1-x}$N
and  In$_x$Ga$_{1-x}$N, respectively, and the solid lines are obtained  by
fitting the results with Eq. (\ref{uno}).

After  comparing our results with the experimental and theoretical
works found in literature for both alloys we conclude that: 1) Our
results  for  the  binary  compounds are in  agreement  with
the published data, both experimental  and theoretical.  
2) For  the Al$_x$Ga$_{1-x}$N alloy, the structural parameters calculated  in
this work are in agreement with those reported in Ref. \cite{SYoshida}.
For  this  alloy,  the bowing parameters for $a,\ c,\ {\rm and}\ u$,  have
positive  values,  indicating  a  downward  bowing.  This  is   in
agreement  with  experimental  and  theoretical  results  reported
previously.  3) For the In$_x$Ga$_{1-x}$N alloy, the bowing  parameter  of
the $a\ {\rm and}\ c$ lattice constants have a negative value, while  the
internal  parameter, $u$, has a positive value. Reported theoretical
calculations do not give any information about the bowing for  the
$u$  parameter and it is not possible to obtain it from experimental
measurements. To the best of our knowledge, this is the first time
that the value of this parameter has been reported.

\subsection{Electronic Structure for the Al$_x$Ga$_{1-x}$N and 
In$_x$Ga$_{1-x}$N alloys}

Before presenting our results of the electronic structure, we
provide a summary of some representative theoretical and
experimental results reported in literature.

\subsubsection{Al$_x$Ga$_{1-x}$N}

\begin{figure*}[t]%
\includegraphics*[width=\linewidth]{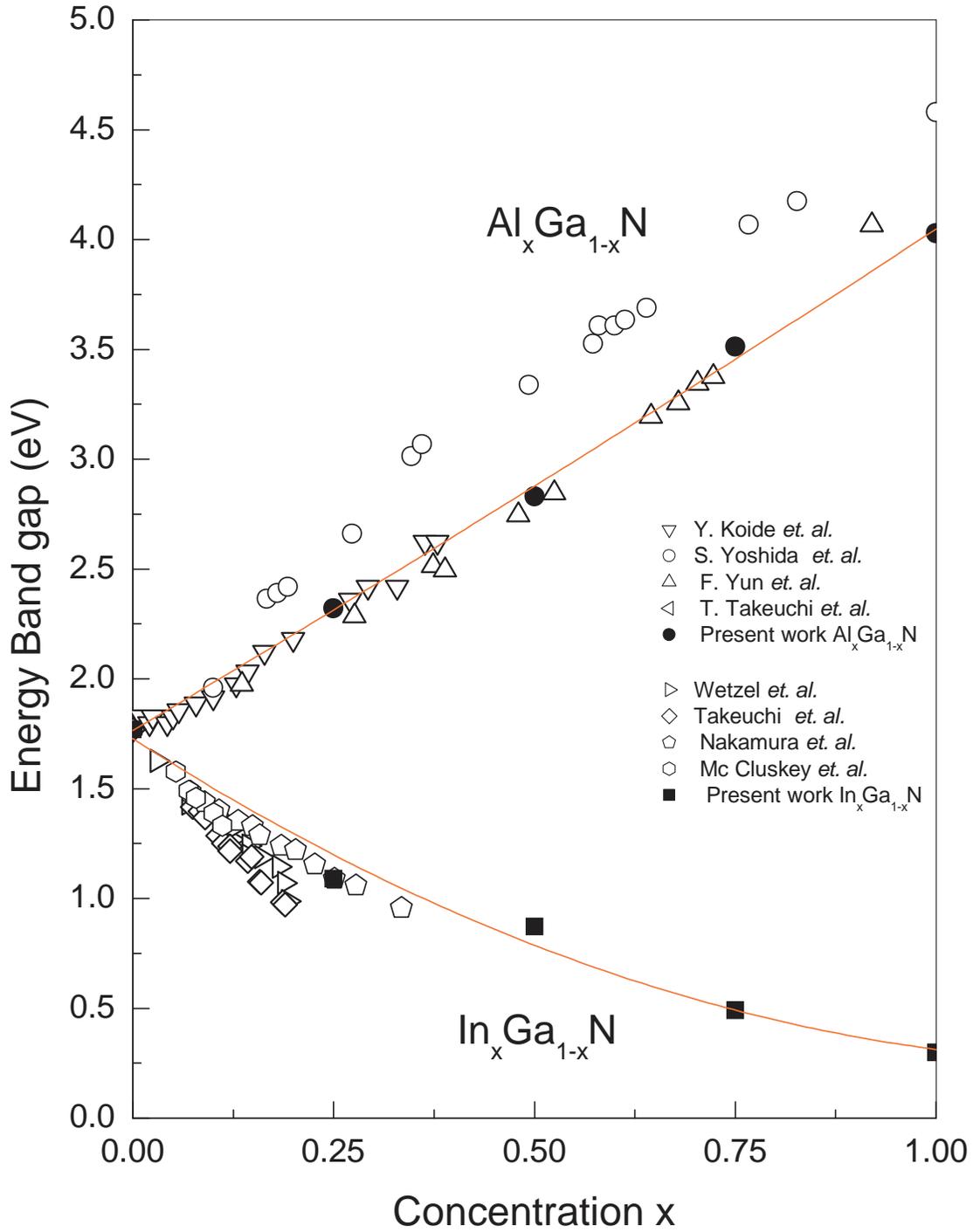}
\vspace{-2cm}
\caption{ Variation of the band gap for the
Al$_x$Ga$_{1-x}$N and
In$_x$Ga$_{1-x}$N alloys as a function of the
concentration, $x$. We present the experimental results from Table 3
and the solid lines represent the proposed adjustment to our results
(solid points)\cite{gapproblem}.}
\label{figura3}
\end{figure*}

Hagan  et al. \cite{JHagan} and Baranov et al. \cite{BBaranov} 
were the  first
to  demonstrate  experimentally the  existence  of  the  Al$_x$Ga$_{1-x}$N
alloy.  Many  other groups have measured its lattice constant, $c$,
the  optical bowing parameter, $\delta$, and the energy gap as a function
of   the  concentration.  The  magnitude  of  the  optical  bowing
parameter  accounts for the deviation of the  band  gap  from  the
linear  dependence.  Using  MBE,  S.  Yoshida  et  al.  \cite{SYoshida}
measured  the  lattice constant and the band gap  for  the  entire
interval of concentrations ($0 < x < 1$). Comparing our results  for
the lattice constant with those reported by this author, there  is
a  very good agreement over the entire interval of concentrations.
Although  our  calculated band gap value  shows  the  experimental
trend  reported  by Yoshida et al. \cite{SYoshida}, we obtain  slightly
different values as can be seen in Fig. \ref{figura3}. However,  other
reported  values  for  the band gap are  well  reproduced  in  our
calculations  (see  Fig. \ref{figura3}). There are  many  experimental
reports  for this system obtained by different techniques and  for
different  values  of the concentration. In all  these  references
there  is  not  a general agreement concerning if the  fundamental
parameters,  i.e. the lattice constants and the  band  gap  value,
follow  Vegard's law. Positive, negative, or small values  of  the
optical  bowing  parameter can be found throughout the  literature
\cite{YKoide,SYoshida,DKWickenden,KItoh,MAKhan,THTakeuchi}.
Theoretical  results have been reported using the {\bf k$\cdot$p} method,  
the semi--empirical  pseudopotentials method, {\it ab initio} LDA,  DFT--LDA
using  molecular dynamics, and plane waves pseudopotentials  using
DFT--LDA.

\subsubsection{In$_x$Ga$_{1-x}$N}

The first In$_x$Ga$_{1-x}$N alloy with a high degree of ordering in layers
grown  on  sapphire  (0001)  using  metal--organic  chemical  vapor
deposition  (MOCVD) were obtained by Ruterana et al.  \cite{PRuterana}.
Samples  grown  with different concentrations and using  different
techniques have been studied and authors report various values for
the  optical bowing parameter. In Table \ref{tabla3} we summarize  the
theoretical and experimental results obtained from the  literature
for  both alloys along with our calculated values obtained for the
optical bowing parameter. It can be seen from the table that there
is  a  large  discrepancy  between the reported  experimental  and
theoretical  values.  For theoretical calculations,  most  of  the
reported  values  are  greater than one and  they  show  a  larger
scattering of the numerical values, especially for the case of the
In$_x$Ga$_{1-x}$N alloy.

In  Table \ref{tabla4} we show the obtained results for the energy gap
of   the  Al$_x$Ga$_{1-x}$N  and  In$_x$Ga$_{1-x}$N  alloys  
from  our  {\it ab  initio}
calculations for different values of the concentration, $x$.  These
values   correspond  to  $0,\ 25,\ 50,\ 75\ {\rm and}\ 100$ \%  
Al  and   In
substitution, respectively.  The plots corresponding to these data
are  displayed  in Fig. \ref{figura3}. It can be seen  that  when  the
concentration  of  Al  is increased the Al$_x$Ga$_{1-x}$N  alloy  shows  a
nearly  linear dependence. On the other hand, the In$_x$Ga$_{1-x}$N  alloy
shows   a   non--linear  dependence  when  we   increase   the   In
concentration.  For both alloys we fit the gaps  obtained  by  our
calculations  to  a  non--linear  dependence  using  the  quadratic
phenomenological function:
\begin{equation}
E_g(x) = x E_{g,A} + (1-x) E_{g,B} -\delta x(1-x);
\label{dos}
\end{equation}
where  $E_{g,A}$ and E$_{g,B}$ corresponds to the gap  of the AlN  (InN)
and  GaN  for  the  Al$_x$Ga$_{1-x}$N (In$_x$Ga$_{1-x}$N) alloy. 
Substituting  the
values of Table \ref{tabla4} into Eq. (\ref{dos}) we found 
$\delta = 0.3185$ and 0.9990  for
Al$_x$Ga$_{1-x}$N and In$_x$Ga$_{1-x}$N, respectively. 
The solid lines in  Fig.  3
correspond to the non--linear fitting given by Eq. (\ref{dos}). It  can
be  seen  from  this  figure  that there  is  a  clear  non--linear
dependence for the In$_x$Ga$_{1-x}$N alloy.

Concerning  the  electronic  structure  of  these  alloys,   after
comparing  our results with those reported in literature we  conclude
that: 1) For both alloys, most of the experimental results consider
concentrations  lower  than  x=0.5. This  could  account  for  the
scattered  values  reported  for the  bowing  parameter;  however,
experiments  considering the whole interval of concentrations  are
better  reproduced by our calculations. 2) The obtained  value  in
this  work of $\delta = 0.3185$ for the bowing parameter of the 
Al$_x$Ga$_{1-x}$N
alloy   is  in  agreement  with  most  of  the  experimental   and
theoretical  results  reported previously. Although  many  authors
claim  that  the  dependence  on concentration  of  the  band  gap
parameter of this alloy should follow a Vegard's--type law 
($\delta = 0.0$),  
it has become more accepted that there is a small deviation in
the  linear  dependence. 3) For the In$_x$Ga$_{1-x}$N alloy  the  reported
results  show  a large scattering, especially in the  experimental
data.  This has been partially explained in the literature as  due
to  an  inaccurate determination of the concentration. The quality
of  the  samples  and  the  measurement technique  also  plays  an
important  role  in  the  determination  of  the  optical   bowing
parameter.  For  this  alloy, the scattering  in  the  theoretical
results  reported  in  the literature is lower.  Our  calculations
determined a value of $\delta = 0.9990$, which is in good agreement  with
those reported previously.

Figure  \ref{figura3}  shows a comparison of our theoretical results with  the
reported  experimental  results for  the  band  gap  energy  as  a
function of concentration for the Al$_x$Ga$_{1-x}$N and In$_x$Ga$_{1-x}$N  alloys.
It  can  be seen from the figure that there is excellent agreement
between our results and those reported, especially for the 
Al$_x$Ga$_{1-x}$N  alloy.  
Most  of  the reported results for the  In$_x$Ga$_{1-x}$N
alloy  are  given  for  low concentrations  of  the  In  impurity.
Nevertheless,   our  theoretical  results  follow   the   tendency
demonstrated  by  experimental  and theoretical  results  reported
previously.

\subsection{Calculated effective masses}

As  a by--product of our electronic band structure calculations, it
is  easy to compute the curvature of the minimum of the conduction
band as well as the maximum of the valence band in the vicinity of
the  $\Gamma-$point. From these values the effective  masses  of  the
electrons and holes can be obtained.

At the $\Gamma-$point the $s-$like conduction band effective mass can
be obtained through a simple parabolic fit using the definition of
the effective mass as the second derivative of the energy band
with respect to the wave vector, {\bf k}, via:
\begin{equation}
\frac{m_o}{m^*} = - \frac{m_o}{\hbar^2}\frac{\partial^2 E}{\partial k^2}
\label{tres}
\end{equation}
where  $m^*$ is the conduction electron effective mass and $m_o$ is  the
free electron mass.

The  valence  band states at the $\Gamma-$point are derived  from  
$p-$bonding states and for the wurtzite crystals these states are  not
as symmetric as the conduction band. However, we can calculate the
curvature   of  the  valence  band  maximum  using  the  following
approach: if the spin--orbit interaction were neglected, the top of
the  valence  band would have a parabolic behavior.  This  implies
that  the  highest valence bands are parabolic in the vicinity  of
the  $\Gamma-$point.  In this work, all the systems  studied  satisfy
this parabolic condition of the maximum of the valence band at the
$\Gamma-$point  \cite{Edmundo}. Within this approach,  and  by  using  the
appropriate  expression of Eq. (\ref{tres}) (using a plus sign  instead
of  the  minus  sign  in the prefactor), we  have  calculated  the
effective masses of the heavy holes at the $\Gamma-$point.

Table \ref{tablamasas} shows our calculated effective masses for the binary
compounds, AlN, GaN, and InN, as well as for their related alloys.
This  table also includes theoretical and experimental values that
were reported in literature, for comparison.

For   the  binary  compounds,  we  conclude  that  our  calculated
effective  masses  are  in  the range  of  most  of  the  reported
theoretical and experimental values. Moreover, we obtain that  our
calculated   values  are  in  good  agreement  with   the   values
recommended by Vurgaftman and Meyer \cite{IVurgaftman}.

We  report  in the same table our calculated effective masses  for
the  alloys  studied in this work. The calculated  value  for  the
conduction  electron effective mass for Al$_x$Ga$_{1-x}$N increases  as  a
function of the Al concentration. It should be noted that although
we  could  not  find  published data for  this  parameter  in  the
wurtzite  phase,  we did find calculated values for  the  electron
effective  mass  of  the alloy in the cubic  phase  \cite{RPaiva}.  The
behavior for our calculations and the reported cubic phase  values
show the same trend. The same behavior was  also  obtained
for the heavy hole effective mass.

The calculated effective masses for In$_x$Ga$_{1-x}$N show a non monotonic
behavior  as a function of the In concentration as seen  in  Table
\ref{tablamasas}.  Comparison  with empirical pseudopotential  calculations
for  the conduction electron effective mass in the cubic phase  of
this alloy \cite{KKassali}, reported for intermediate values of the  In
impurity, shows that we obtain similar behavior for our calculated
effective masses.

\subsection{Zone center phonon calculation}

\begin{figure*}[t]%
\includegraphics*[width=\linewidth]{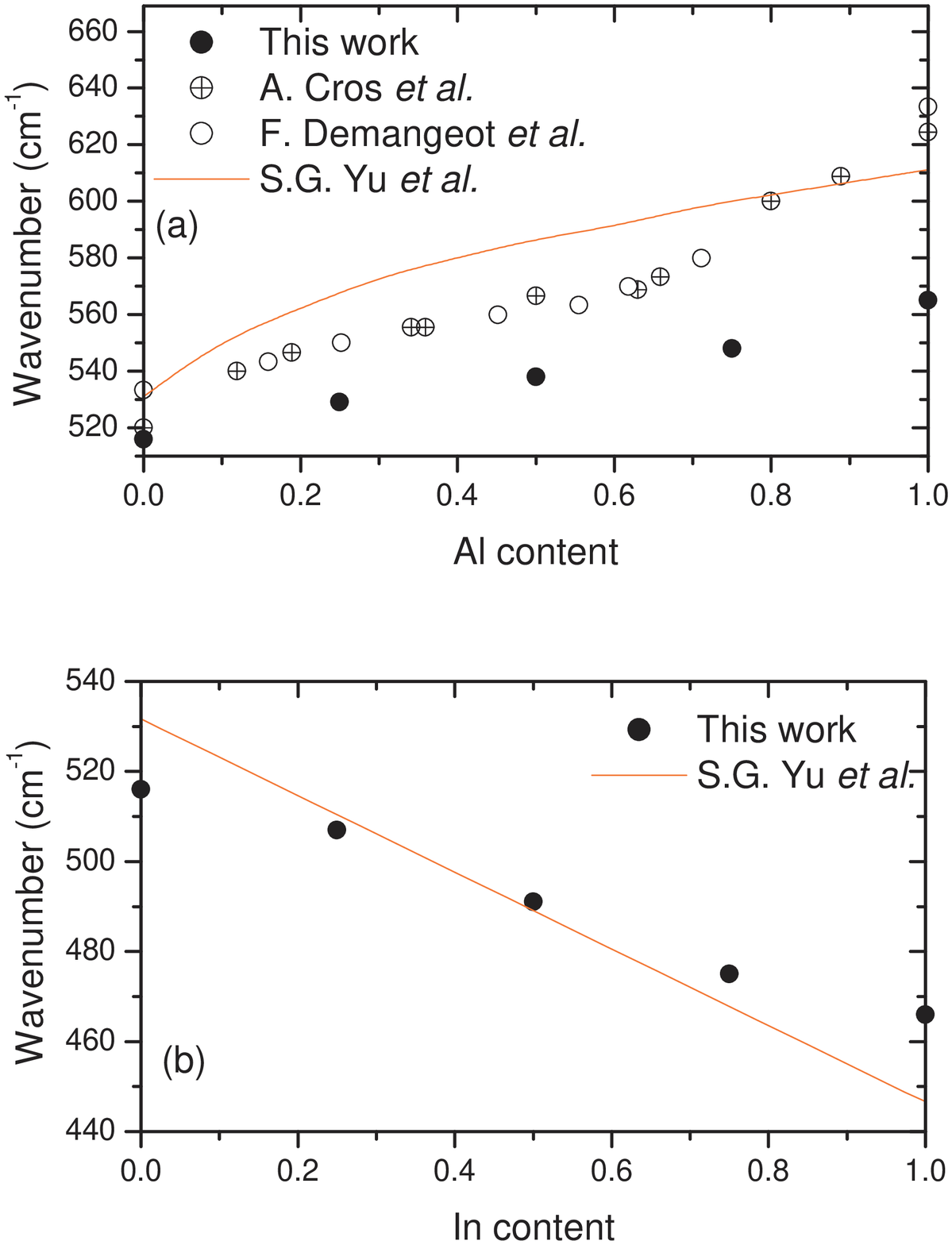}
\vspace{-2cm}
\caption{$A_1(TO)$ phonon dependence with Al concentration for 
Al$_x$Ga$_{1-x}$N alloy (upper figure),
and as a function of the In 
concentration for In$_x$Ga$_{1-x}$N
(lower figure). 
We present our calculations (full circles), 
the experimental results of 
Ref. \cite{ACros} (open circles with plus),
and Ref. \cite{Demangeot} (open circles).
Solid line depicts the calculated values from 
Ref. \cite{SGYu}.}
\label{figuraphn}
\end{figure*}

Using  the frozen phonon approach, we can compute the zone  center
phonon modes, A$_1(TO)$ and E$_1(TO)$, by considering the change of  the
total  energy as a function of the displacement, $u$, of the atoms
in  the unit cell from their equilibrium positions. In particular,
we  can compute the A$_1(TO)$ mode by considering small displacements
of  the  atoms along the optical axis (the c--axis) of the wurtzite
phase.  Then,  as we are at the minimum of the total  energy of
the system, the
perturbation  around  this  minimum allows  us  to  use  the  next
parabolic  approach  to the total energy  as  a  function  of  the
displacement:
\begin{equation}
E_{Total} (u) = E_o + u E_1 + u^2 E_2  = m \omega;
\end{equation}
Where $E_i$ ($i = 0,\ 1,\ 2$) are fit parameters, $\mu$ is the reduced mass,
and  $\omega$   is the frequency. With this approach, our  calculated
value  for  the  $A_1(TO)$ mode for the binary compounds  shows  good
agreement with most of the experimental reports as shown in  Table
\ref{tablaphn}. 
Figure \ref{figuraphn}a shows our calculated A$_1(TO)$ mode for
the  Al$_x$Ga$_{1-x}$N alloy, which is compared with the experimental data
of   Refs.   \cite{ACros,Demangeot}, as well as the theoretical
calculations of Ref. \cite{SGYu}.  
Although,  we  obtain  slightly
different  values over the range of Al concentration,  we  observe
that  the  figure depicts the same trend shown by the experimental
reports.  Figure \ref{figuraphn}b shows our calculated A$_1(TO)$  mode  for
the   In$_x$Ga$_{1-x}$N   alloy  and compares  them   with   theoretical
calculations reported in Ref. \cite{SGYu}. As in the previous case,  it
can  be  seen  that  our calculated values  for  the  A$_1(TO)$  mode
reproduce the trend reported in Ref. \cite{SGYu}.
In this way we show  that  our
calculations properly reproduce the reported values of the  A$_1(TO)$
mode  for the alloys and their binary parent compounds studied  in
this work.

\section{Conclusions}

We  have  calculated the structural and electronic  properties  of
wurtzite  AlN,  GaN, InN, and their related alloys, Al$_x$Ga$_{1-x}$N  and
In$_x$Ga$_{1-x}$N.  We  found  that,  for both  alloys,  their  structural
parameters  as a function of the concentration, $x$, do  not  follow
Vegard's law. We observe that for the Al$_x$Ga$_{1-x}$N alloy, the  
$a,\  c,\ {\rm and}\  u$ parameters have a positive bowing parameter, 
of which,  the
lattice constant, $c$, demonstrates the largest value. On the  other
hand, when the concentration increases in the In$_x$Ga$_{1-x}$N alloy, the
bowing parameter for the $a$ and $c$ lattice constants have a negative
value  and  the  bowing parameter for the internal  parameter, $u$,
remains  positive.  From  our  results  of  the  electronic   band
structure  calculations  we  obtained the  band  gap  energy  as  a
function  of the concentration, $x$, and characterized the deviation
from  the linear behavior by calculating 
the optical bowing parameter, $\delta$.  
We obtain a small optical bowing parameter $\delta = 0.3185$ 
for the
Al$_x$Ga$_{1-x}$N  alloy,  which  is  consistent  with  results   reported
previously. The deviation from the linear behavior is more drastic
for  the In$_x$Ga$_{1-x}$N alloy, which has $\delta = 0.9990$
in accordance  with
most  of  the  experimental  results  found  in  literature.   The
effective  masses  of  the  systems  studied  in  this  work  were
calculated  and we found that our calculated effective masses  for
the  binary compounds, AlN, GaN, and InN, are in the range of  the
reported  values  in the recent literature. 
To  the  best  of  our
knowledge,  this  is the first time that the deviation  parameter
for the internal parameter, $\epsilon_u$, as well as the effective masses
for  these  alloys have been reported. Finally, using  the  frozen
phonon approach we have computed the A$_1(TO)$ mode for the different
systems studied in this work. From our calculations we have  found
good  agreement with experimental values reported for  the  binary
compounds.  For  the  ternary alloys, our  calculations  reproduce
experimental   values  for  Al$_x$Ga$_{1-x}$N,  as  well  as   theoretical
predictions for In$_x$Ga$_{1-x}$N.

\begin{acknowledgments}
This work was done using the Computer Facilities of 
IPICYT, San Luis Potos\'\i, M\'exico.
This work was partially supported by VIEP--BUAP under grant
VIEP--BUAP 1/I/EXC/05, VIEP--BUAP 01/EXC/06--I.
ELA acknowledges to CONACyT under grat No.165404.
This work was 
completed while DO was in a research stay 
at Freie Universit\"at Berlin, the hospitality of 
H. Kleinert is greatly appreciated. DO 
gratefully acknowledges financial support 
from CINVESTAV--IPN and CONACYT--M\'exico.
\end{acknowledgments}

\newpage

\begin{table*}[h]
\caption{\label{tabla1} Structural parameters (in \AA) for the
AlN, GaN, and Al$_{x}$Ga$_{1-x}$N alloys.
The lattice parameters $a$ and $c$ are given in \AA.}
\begin{tabular}{p{2cm}p{1cm}p{2cm}p{4cm}p{4cm}}
\hline
       &  & Present Work & Exp. Results          &  Other Calc. \\
\hline
GaN    & a & 3.2209 
            & 3.1890\footnote{Ref. \cite{SStrite}}, 
                     3.1892\footnote{Ref. \cite{TDetchprohm}}
             & 3.1660\footnote{Ref. \cite{ZDridi}}, 
                      3.189\footnote{Ref. \cite{AFWright}}, 
                           3.2\footnote{Ref. \cite{CBungaro}} \\
              & &       
               & 3.1880\footnote{Ref. \cite{MLeszczynski}}, 
                        3.19\footnote{Ref. \cite{HSchulz}}
                &3.1800\footnote{Ref. \cite{JSerrano}}, 
                       3.1986\footnote{Ref. \cite{AZoroddu}}, 
                             3.17\footnote{Ref. \cite{PCarrier}}, 
                    3.183\footnote{Ref. \cite{Liou}} \\
       & c & 5.2368 
            & 5.1850$^{a}$, 5.185$^{b}$
             & 5.1540$^{c}$, 5.185$^{d}$, 5.2200$^{e}$ \\
              & &       
                & 5.18561$^{f}$, 5.189$^{g}$
                 & 5.1898$^{h}$, 5.2262$^{i}$ 5.151$^{j}$,
                    5.178$^k$ \\
      & u & 0.3780 
           & 0.3768$^{a}$, 0.377$^{g}$
            & 0.3770$^{c}$, 0.3768$^{d}$, 0.3760$^{e}$ \\
              & & & & 0.3760$^{h}$, 0.3772$^{i}$, 0.3768$^{j}$\\\hline
Al$_{0.25}$Ga$_{0.75}$N  &a& 3.2059 & See Ref. \cite{SYoshida,FYun} & 
                                      3.163$^k$ \\
                   &c& 5.1338 &  & 5.137$^k$ \\
                   &u& 0.3781 &  &\\\hline
Al$_{0.50}$Ga$_{0.50}$N  &a& 3.1719 & See Ref. \cite{SYoshida,FYun} & 
                                      3.139$^k$\\
                   &c& 5.1012 &  & 5.085$^k$\\
                     &u & 0.3790 &  &\\\hline
Al$_{0.75}$Ga$_{0.25}$N  &a& 3.1601 &  & 3.098$^k$ \\
                   &c& 5.0870 & See Ref. \cite{SYoshida,FYun} & 4.990$^k$ \\
                   &u& 0.3791 &  &\\\hline
AlN         & a & 3.1411 
                 & 3.1120$^{a}$, 3.11$^{g}$
                  & 3.0920$^{c}$, 3.084$^{d}$, 3.1$^{e}$ \\
                 & & & 3.1106\footnote{Ref. \cite{MTanaka}}
                  & 3.0610$^{h}$, 3.10954$^{i}$, 3.098$^{j}$, 3.076$^k$ \\
              & c & 5.0268 
                   & 4.9820$^{a}$, 4.98$^{g}$
                    & 4.954$^{c}$, 4.9948$^{d}$, 5.010$^{e}$ \\
                   & & & 4.9795$^l$
                    & 4.8976$^{h}$, 4.9939$^{i}$, 4.9599$^{j}$, 4.935$^k$\\
              & u & 0.3805 
                   & 0.3819$^{a}$, 0.3821$^{g}$
                    & 0.3821$^{c}$, 0.3814$^{d}$, 0.3800$^{e}$ \\
                   & & & & 0.3820$^{h}$, 0.3819$^{i}$, 0.3819$^{j}$\\
\hline \\
\end{tabular}
\end{table*}

\begin{table*}[t]
\caption{\label{tabla2} Structural parameters (in \AA) for the
 GaN, InN and In$_{x}$Ga$_{1-x}$N alloys.
The lattice parameters $a$ and $c$ are given in \AA.}
\begin{tabular}{p{2cm}p{1cm}p{2cm}p{4cm}p{4cm}}
\hline
         &  & Present Work & Exp. Results & Other Calc.  \\
\hline
GaN  & a & 3.2209 
          & 3.1890\footnote{Ref. \cite{SStrite}}, 
                   3.1892\footnote{Ref. \cite{TDetchprohm}}
           & 3.1660\footnote{Ref. \cite{ZDridi}}, 
                   3.189\footnote{Ref. \cite{AFWright}}, 
                        3.2000\footnote{Ref. \cite{CBungaro}} \\
          & & & 3.1880\footnote{Ref. \cite{MLeszczynski}}, 
                      3.19\footnote{Ref. \cite{HSchulz}}
             & 3.1800\footnote{Ref. \cite{JSerrano}}, 
                     3.1986\footnote{Ref. \cite{AZoroddu}}, 
                           3.17\footnote{Ref. \cite{PCarrier}} \\
     & c & 5.2368 
          & 5.1850$^{a}$, 5.185$^{b}$
           & 5.1540$^{c}$, 5.185$^{d}$, 5.2200$^{e}$ \\
          & &       
             & 5.18561$^{f}$, 5.189$^{g}$
              & 5.1898$^{h}$, 5.2262$^{i}$, 5.151$^{j}$ \\
     & u & 0.3780
          & 0.3768$^{a}$, 0.377$^{g}$
           & 0.3770$^{c}$, 0.3768$^{d}$, 0.3760$^{e}$ \\
              & & & &0.3760$^{h}$, 0.3772$^{i}$, 0.3768$^{j}$ \\\hline
In$_{0.25}$Ga$_{0.75}$N  &a & 3.3298 &   &\\
                     &c & 5.3987 &   &\\
                     &u & 0.3791 &   &\\
\hline
In$_{0.50}$Ga$_{0.50}$N  &a & 3.4128 &   &\\
                     &c & 5.5257 &   &\\
                     &u & 0.3792 &   &\\
\hline
In$_{0.75}$Ga$_{0.25}$N  &a & 3.4969 &   &\\
                     &c & 5.6333 &   &\\
                     &u & 0.3796 &   &\\
\hline
InN & a & 3.5440 
         & 3.5365\footnote{Ref. \cite{Davydov}}, 
                  3.5378\footnote{Ref. \cite{Paszkowicz}}
          & 3.520$^{c}$, 3.501$^{d}$, 3.480$^{e}$ \\
         & & & 3.548\footnote{Ref. \cite{Tansley}}, 
                    3.540\footnote{Ref. \cite{Kubota}}
          & 3.525$^{h}$, 3.614$^{i}$, 3.546$^{j}$ \\ 
    & c & 5.7228 
         & 5.7039$^{k}$, 5.7033$^{l}$
          & 5.675$^{c}$, 5.669$^{d}$, 5.64$^{e}$ \\
         &  & & 5.76$^{m}$, 5.705$^{n}$
           & 5.68583$^{h}$, 5.8836$^{i}$, 5.7162$^{j}$ \\
    & u & 0.3806 
         & 
          & 0.3799$^{c}$, 0.3784$^{d}$, 0.378$^{e}$\\
         &  & &        
          & 0.379$^{h}$, 0.37929$^{i}$, 0.379$^{j}$\\ \hline
\end{tabular}
\end{table*}

\begin{table*}[h]
\caption{\label{tabla3} Experimental and theoretical results for
the optical bowing parameter $\delta$ for Al$_{x}$Ga$_{1-x}$N and
In$_{\textrm{x}}$Ga$_{\textrm{1-x}}$N.} {\small
\begin{tabular}{p{7cm}p{1.2cm}p{7cm}p{1.5cm}} 
\hline
Al$_x$Ga$_{1-x}$N & & & \\
\hline
Experimental & ${\delta}$ [eV]  & Theoretical & $\delta$ [eV] \\
\hline  
\cite{YKoide} Y. Koide, for $0 \le x \le 0.4$ using MOVPE &
$\approx 1$ & 
\cite{MGoano} M. Goano  using pseudopotentials and VCA &0.069 \\
\cite{SYoshida} S. Yoshida, $0 \le x \le 1$ using  MBE at $700^{\circ}$ &
$\approx 0$ &
\cite{ZDridi} Z. Dridi using FP--LAPW LDA and VCA & 0.710 \\
\cite{FYun} F. Yun, for $0 \le x \le 1$ using MBE at $600-670^{\circ}$
& 1 & 
\cite{DKWickenden} D. K. Wickenden,$0 \le x \le 0.4$ using MOCVD & $\approx 0$\\
\cite{MAKhan} M. A. Khan, $0 \le x \le 1$ using MOCVD at
$915^{\circ}$ & $\approx 0$ & 
\cite{KChen} K. Chen using {\it ab initio} Molecular dynamics& 1.40\\
\cite{THTakeuchi} T. Takeuchi, for $0 \le x \le 0.25$ using MOVPE &$\approx 1$
& 
\cite{SKPugh} S. K. Pugh using first principles
${\bm k}\cdot {\bm p}$ method &$\approx 0$\\
\cite{OAmbacher} O. Ambacher Review article &1         &Value
calculated in this work& 0.3185\\
\hline
In$_x$Ga$_{1-x}$N & & & \\
\hline
Experimental & ${\delta}$ [eV]  & Theoretical & $\delta$ [eV] \\
\hline 
\cite{Davydov} V. Yu. Davydov {\it et al.} for $0.36 < x < 1$ & 2.50    & 
\cite{MGoano} M. Goano {\it et al.} &1.05\\
\cite{THTakeuchi} T. Takeuchi {\it et al.} &3.20    &
\cite{ZDridi} Z. Dridi {\it et al.} &1.70\\
\cite{SNakamura} S. Nakamura {\it et al.}, for $0\le x\le 0.4$ & 1.00 & 
\cite{CCaetano} C. Caetano {\it et al.}, using FP and LDA &1.44\\
\cite{MDMcCluskey} M. D. McCluskey {\it et al.}, for $0\le x \le 0.12$&3.5
&
\cite{AFWright} A. F. Wright {\it et al.}&0.1677\\
\cite{CWetzel} C. Wetzel {\it et al.}, for $0\le x \ge 0.2$ using MOVPE &2.6   
& Value calculated in this work &0.999\\
\cite{JWu} J. Wu {\it et al.}, for $0\le x\le 0.5$ using MBE & 1.4  &  & \\
\hline
\end{tabular}}
\end{table*}

\newpage

\begin{table*}[h]
\caption{\label{tabla4} Band gap energy (eV) for the
Al$_{\textrm{x}}$Ga$_{\textrm{1-x}}$N and
In$_{\textrm{x}}$Ga$_{\textrm{1-x}}$N alloys as a function of the
concentration, $x$, obtained in the present work.}
\begin{tabular}{p{5cm}{c}p{5cm}{c}}

\hline
System           &E$_{gap}$(eV)   \\\hline
GaN                 &1.768\\
Al$_{0.25}$Ga$_{0.75}$N &2.319\\
Al$_{0.50}$Ga$_{0.50}$N &2.830\\
Al$_{0.75}$Ga$_{0.25}$N &3.5123\\
AlN                 &4.027\\
\hline
GaN                 &1.768\\
In$_{0.25}$Ga$_{0.75}$N &1.088\\
In$_{0.50}$Ga$_{0.50}$N &0.871\\
In$_{0.75}$Ga$_{0.25}$N &0.489\\
InN                 &0.299\\
\hline
\end{tabular}
\end{table*}

\begin{table*}[ht]
\caption{\label{tablamasas} The electron 
and hole effective masses for AlN, GaN, InN and their alloys. 
$m^{\ast}$ denote  the average masses. 
The average effective mass can be obtained using 
$m^{\ast}=[m^{\perp}_{\Gamma \to M} m^{\perp}_{\Gamma \to K} 
m^{\parallel}_{\Gamma \to A}]^{1/3}$.
Where, $m^{\perp}$ and $m^{\parallel}$ 
denote the {\bf k} direccion dependent masses 
perpendicular and parallel to the c axis, respectively. 
All values are in units of a free--electron mass $m_{0}$.
For a recommended set of values for the binary compounds see
Ref. \cite{IVurgaftman}.}
\begin{tabular}{l|c|c|c|ccc|c}
\hline
     &  &\multicolumn{2}{c}{Present work}&\multicolumn{4}{c}{Other results}\\
\hline
     & & $m_{e}$ & $m_{h}$ & \multicolumn{3}{c}{$m_e$} 
                                             & \multicolumn{1}{c}{$m_{h}$} \\
\hline
AlN  & $m^{\perp}$ & 0.3012 & 4.3243 
                              &  0.30\footnote{Ref. \cite{IVurgaftman}} 
                                   & 0.33\footnote{Ref. \cite{Rinke}} 
                                     & 0.33\footnote{Ref. \cite{PCarrier}}
                                       & 4.35$^{c}$ \\
     & $m^{\parallel}$ & 0.2847 & 0.2427 & 0.32$^a$ & 0.32$^b$ & 0.32$^c$ 
                                        & 0.28$^c$ \\
     & $m^{\ast}$      & 0.2956 & 1.6528 & 0.31\footnote{Ref. \cite{JDAlbrect}} 
                                        & 0.48\footnote{Ref. \cite{BEFoutza}}
                                           & & \\
\hline
Al$_{0.75}$Ga$_{0.25}$N & $m^{\perp}$   &0.2682& 3.5247 & & & & \\
                    &$m^{\parallel}$&0.2913& 0.9325 & & & & \\
                    &$m^{\ast}$     &0.2749& 1.4743 & & & & \\
\hline
Al$_{0.50}$Ga$_{0.50}$N &$m^{\perp}$    &0.2330& 2.3039 & & & & \\
                    &$m^{\parallel}$&0.2411& 0.5345 & & & & \\
                    & $m^{\ast}$    &0.2427& 1.2961 & & & & \\
\hline
Al$_{0.25}$Ga$_{0.75}$N &$m^{\perp}$    &0.2020& 3.1232 & & & & \\
                    &$m^{\parallel}$&0.1958& 0.1741 & & & & \\
                    &$m^{\ast}$     &0.2000& 1.1969 & & & & \\
\hline
GaN  & $m^{\perp}$  & 0.1491 & 2.1072 & 0.20$^{a}$ 
                                        & 0.21$^{b}$ 
                                         & 0.22$^{c}$ 
                                           & 0.39$^{c}$ \\
     & $m^{\parallel}$ & 0.1803 & 2.1048 
                                  & 0.20$^{a}$
                                    & 0.19$^{b}$ & 0.20$^{c}$ 
                                                       & 2.04$^{c}$ \\
     & $m^*$ & 0.1692 & 2.1412 & 0.22\footnote{Ref. \cite{PPerlin}}
                                 & 0.23\footnote{Ref. \cite{YJWang}}
                                   & 0.20\footnote{Ref. \cite{SKOLeary}} 
                                     & 2.2\footnote{Ref. \cite{JSIm}}\\
     &  &  &  & 0.18\footnote{Ref. \cite{SElhamri}}
                & 0.20\footnote{Ref. \cite{BEFoutza}}
                  & 0.24\footnote{Ref. \cite{TYLim}} & \\
    & & & & 0.215\footnote{Ref. \cite{ASaxler}} & & & \\
\hline
In$_{0.25}$Ga$_{0.75}$N &$m^{\perp}$    &0.1069& 2.5409&&&&\\
                    &$m^{\parallel}$&0.0968& 2.4447&&&&\\
                    &$m^{\ast}$     &0.1035& 2.4953&&&&\\
\hline
In$_{0.50}$Ga$_{0.50}$N &$m^{\perp}$    &0.1182& 2.7958&&&&\\
                    &$m^{\parallel}$&0.0921& 2.6116&&&&\\
                    &$m^{\ast}$     &0.1025& 2.8371&&&&\\
\hline
In$_{0.75}$Ga$_{0.25}$N &$m^{\perp}$    &0.0717& 2.0304&&&&\\
                   &$m^{\parallel}$&0.0623& 2.1486&&&&\\
                    &$m^{\ast}$     &0.0781& 2.2179&&&&\\
\hline
InN & $m^{\perp}$ & 0.1299 & 1.9096 & 0.07$^{a}$ & 0.068$^{b}$ 
                                     & 0.07$^{c}$ 
                                      & 2.967\footnote{Ref. \cite{Fritsch}} \\
                    &               &      &       &0.068$^{n}$ & & \\
    & $m^{\parallel}$ & 0.0892 & 2.2051 & 0.07$^{a}$ & 0.065$^{b}$ 
                                         & 0.06$^{c}$ 
                                          & 2.566$^{n}$ \\
                    &               &      &       &0.072$^{n}$ & & \\
   & $m^{\ast}$ & 0.1146 & 2.0301 & & & & \\
\hline
\end{tabular}
\end{table*}

\begin{table}[ht]
\caption{\label{tablaphn} 
Calculated $A_1(TO)$ mode for the different systems studied in this 
work. Comparison with other calculations and experimental data
(all values in cm$^{-1}$).}
\begin{tabular}{l|c|ccc|ccc}
\hline
\hline
System & Present work &\multicolumn{3}{c}{Other results} & 
                                            \multicolumn{3}{c}{Experimental}\\
\hline
\hline
GaN   & 516 & 537\footnote{Ref. \cite{Gorczyca}} 
              & 545\footnote{Ref. \cite{Aouas}} & 
                 & 533.8\footnote{Ref. \cite{YuDavidov}} 
                   & 533.5\footnote{Ref. \cite{Zhang}} & \\
\hline
Al$_{25}$Ga$_{75}$N & 529 & &&&&&\\
\hline
Al$_{50}$Ga$_{50}$N & 538 & &&&&&\\
\hline
Al$_{25}$Ga$_{25}$N & 548 & &&&&&\\
\hline
AlN                 & 565 & 649$^a$ & 615$^c$& 619\footnote{Ref. \cite{Karch}} 
                      & 613.8$^c$&&\\
\hline
In$_{25}$Ga$_{75}$N & 507 & &&&&&\\
\hline
In$_{50}$Ga$_{50}$N & 491 & &&&&&\\
\hline
In$_{75}$Ga$_{25}$N & 475 & &&&&&\\
\hline
InN                 & 466 & 443\footnote{Ref. \cite{Kaczmarczyk}} & & 
                      & 447\footnote{Ref. \cite{YuDavidov2}}& 440$^f$ & \\
\hline
\hline
\end{tabular}
\end{table}

\newpage

\newpage

\newpage

\newpage

\end{document}